\documentstyle[12pt,rotate,cite,epsfig]{article}
\input psfig.sty

\newcommand{\beq}{\begin{equation}}
\newcommand{\eeq}{\end{equation}}
\newcommand{\bea}{\begin{eqnarray}}
\newcommand{\eea}{\end{eqnarray}}
\newcommand{\bd}{\begin{displaymath}}
\newcommand{\ed}{\end{displaymath}}

\newcommand{\gm}{\gamma}

\newcommand{\Bg}{B\rightarrow\rho\gm}
\newcommand{\bg}{B\to\rho}
\newcommand{\F}{F^{\bg}(0)}

\newcommand{\es}{\epsilon}

\begin{document}
\bibliographystyle{physics}
\renewcommand{\thefootnote}{\fnsymbol{footnote}}

\author{
Tao Huang${}^{1,2}$~~~Zuohong Li${}^3$
~~~and~~~ Haidong Zhang${}^2$  
\footnote{Email: huangt@hptc5.ihep.ac.cn, zhanghd@hptc5.ihep.ac.cn,
lizhh@ibm320h.phy.pku.edu.cn}\\
{\small\sl ${}^{1}$ CCAST (World Laboratory), P.O.Box 8730, Beijing 
100080, China}\\
{\small\sl ${}^{2}$ Institute of High Energy Physics, Academia Sinica,
P.O.Box 918(4), Beijing 100039, China\thanks{Mailing address} }\\
{\small\sl ${}^{3}$ Department of Physics, Peking University, Beijing 
100871, China}\\
}
\date{}
\title{
{\large\sf
\rightline{BIHEP-Th/98-9}
}
\vspace{1cm}
{\LARGE\sf A New Estimate of $B \rightarrow (\rho, \omega)+ \gamma$ 
In The Light-Cone QCD Sum Rule   }}

\maketitle
\thispagestyle{empty}
\begin{abstract}
\noindent
\par We propose a new estimate of the process
$B \rightarrow (\rho, \omega)+ \gamma$ by using the light-cone
QCD sum rule. The aim is to choose the chiral quark current
operators in order to eliminate the uncertainty arising
from the chiral-even operators in the correlator.
In the case of neglecting the corrections from $\rho$-meson mass, the 
sum rules for form factors are obtained, which are valid to the 
twist-three accuracy.
The resulting branching ratios are consistent 
with the experimental data.

\end{abstract}

\newpage
\setcounter{page}{1}

\setcounter{footnote}{0}
\renewcommand{\thefootnote}{\arabic{footnote}}

\par The rare decays of B mesons have been studied for years for a precise
understanding of the physics of CP-Violation related to the structure of the
CKM mixing matrix elements and testing of standard model to search for new
physics. At present, only upper bounds on the exclusive modes 
$B \rightarrow (\rho, \omega)+ \gamma$ are
available from CLEO$\amalg$, namely $Br(B^0\to\rho^0\gm)\leq 3.9\times10^{-5}$,
$Br(B^0\to\omega\gm)\leq 1.3\times10^{-5}$, and
$Br(B^-\to\rho^-\gm)\leq 1.1\times10^{-5}$ in Ref.[1]. The rare decay
modes $B \rightarrow (\rho, \omega)+ \gamma$ are
very suitable for study with the light-cone QCD sum rule method, which is
developed in Ref.[2,3] recently, and have been investigated in Refs.[4,5]. 
The points of the light-cone QCD sum rule are combining the
traditional QCD sum rule method with the hadronic wavefunction
descriptions of the hard exclusive process. However, instead of the small
distance $x\simeq0$ in the traditional QCD sum rule method,
the operator product
expansion (OPE) in the light-cone sum rule is carried out in terms of
nonlocal operators near the light cone $x^2\simeq0$; and instead of the
vacuum condensate, the nonperturbative dynamics is parametrized as so-called
light-cone wavefunctions classified by their twist. Using the light-cone
QCD sum rule, the authors in Ref.[4] found the form factor
$\F=0.285\pm15\%$ with the b-quark pole mass $m_b=4.8 GeV$ and a
threshold $s_0=33.5\pm0.5 GeV^2$. However, $g^\nu_\perp$ and $g^a_\perp$,
the nonleading
twist distribution wavefunctions of $\rho$ meson used in Ref.[4,5], are not
known well with a drastic effect on the resulting precision. In
addition, there are the uncertainties in the  b-quark
pole mass and condensate parameters, which is an important input parameter
in the light-cone sum rule for the form factor $\F$. Therefore, a further
precision investigation of this rare process is valuable and necessary.
Ref.[6] has made a new estimate of the process $B\to K^\ast\gm$ by choosing
the chiral current operators in the correlator. Similarly,
through the choices of the chiral current operators in the correlation 
function,
the uncertain nonperturbative quantities can be eliminated effectively from
the respective sum rules for the form factor $\F$ and decay constant $f_B$,
and thus a more precise prediction for the rare process can be made.
\par Following the procedure in Ref.[6], we calculate the rare decay $\Bg$
which is induced by the effective Hamiltonian
  \begin{equation}
  H_{eff}={-4G_F\over\sqrt{2}}\xi_t\sum\limits_{j=1}^{8}C_j(\mu)O_j(\mu)
  \end{equation}
where $\xi_t=V_{tb}V^*_{td}$, $C_i(\mu)$ are Wilson coefficients and 
the apparent
forms of operators $O_i(\mu)$ are discussed in Refs.[7,8]. 
\par In the process $\Bg$, the electromagnetic penguin operator
$O_7=\frac{e}{16\pi^2}m_b\overline{d}\sigma^{\mu\nu}\frac{1+\gm_5}{2}bF_{\mu\nu}$ 
plays the most important role 
and it leads to the effective Hamiltonian responsible to $b\rightarrow d+\gm$:
  \begin{equation}
H_{eff}=Cm_b\epsilon^{\mu}\bar d\sigma_{\mu\nu}(1+\gamma_5)q^{\nu}b,
  \end{equation}
where $\epsilon$ and q are the emitted photon polarization vector and momentum,
respectively; the constant C depends on the CKM matrix elements
$V_{tb}V^\ast_{td}$.
\par Then the amplitude for $B(p+q)\to\rho(p,\eta)+\gm(q,\es)$ becomes:
   \begin{equation}
A(B \to \rho\gamma)=Cm_b\epsilon^{\mu}<\rho(p,\eta)|\bar d
\sigma_{\mu\nu}(1+\gamma_5)q^{\nu}b|B(p+q)>.
\end{equation}
With $\sigma_{\mu\nu}\gamma_5=\frac{i}{2}\epsilon_{\mu\nu
\alpha\beta}\sigma^{\alpha\beta}$,
we can parametrize the hadronic matrix element (at $q^2=0$) in
Eq.(3) as
   \begin{equation}
<\rho(p,\eta)|\bar{d}\sigma_{\mu\nu}(1+\gamma_5)q^{\nu}b|B(p+q)>
                 =(-2i\epsilon_{\mu\nu\alpha\beta}
                 \eta^{\nu}q^{\alpha}p^{\beta}
                 +2p\cdot{q}\eta_{\mu}
                 -2q\cdot{\eta p_{\mu}})\F .
\end{equation}
Thus the decay width $\Gamma(\Bg)$ can be easily written as
  \begin{equation}
\Gamma(\Bg)={\alpha\over32\pi^4}G_F^2\left|V_{tb}\right|^2\left|V^*_{td}\right|^2\left|\F\right|^2C_7(\chi_t,m_b)^2(m_b^2+m_d^2)\frac{(m^2_B-m^2_\rho)^3}{m^3_B}.
  \end{equation}
\par Now we are in a position to calculate the form factor $\F$ in the
light-cone QCD sum rule framework. In order to eliminate the uncertain
nonperturbative quantities $g_\perp^\nu$ and $g_\perp^a$,
we adopt the following correlator of the chiral current operators
 \begin{equation}
F_{\mu}(p,q)=i\int d^{4}xe^{iqx}<\rho(p,\eta)|T\bar{d}(x)
                 \sigma_{\mu\nu}(1+\gamma_5)q^{\nu}b(x)\bar{b}(0)
                 i(1+\gamma_5)u(0)|0>,
\end{equation}  
which is different from Ref.[4,5].
$F_\mu(p,q)$ can further be written as two separate terms:
  \begin{eqnarray}
F_{\mu}(p,q)&=&i\int d^{4}xe^{iqx}<\rho(p,\eta)|T(\bar{d}(x)
                 \sigma_{\mu\nu}(1+\gamma_5)q^{\nu}b(x)
                 \bar{b}(0)iu(0))|0>\nonumber\\
            &+&<\rho(p,\eta)|T(\bar{d}(x)\sigma_{\mu\nu}
                 (1+\gamma_5)q^{\nu}
                 b(x)\bar{b}(0)i\gamma_{5}u(0))|0>.
\end{eqnarray}
Both pseudoscalar resonance states $B^h_p$ and scalar resonance states
$B^h_s$ in the hadronic expression give contributions to the correlation
function:
  \begin{eqnarray}
F_{\mu}(p,q)&=&\frac{<\rho|T(\bar{d}
\sigma_{\mu\nu}(1+\gamma_5)q^{\nu}b)|B><B|\bar{b}i\gm_5u|0>}{m^2_B-(p+q)^2}\nonumber\\
            &+&\sum\limits_{h}\frac{<\rho|T(\bar{d}
\sigma_{\mu\nu}(1+\gamma_5)q^{\nu}b)|B_{p(s)}^h><B_{p(s)}^h|\bar{b}i(1+\gamma_5)u
|0>}{m^2_{B_{p(s)}^h}-(p+q)^2}
\end{eqnarray}
which can be parametrized as the Lorentz-covariant form:
  \begin{equation}
F_{\mu}(p,q)=(-2i\epsilon_{\mu\nu\alpha\beta}
\eta^{\nu}q^{\alpha}p^{\beta}+2p\cdot q\eta_{\mu}
-2q\cdot \eta p_{\mu})F((p+q)^2).
\end{equation}
The invariant amplitude $F((p+q)^2)$ obeys a general dispersion relation in
the momentum squared $(p+q)^2$:
  \begin{equation}
F((p+q)^2)=\int^{\infty}_{m_B}\frac{\rho(s)}{s-(p+q)^2}ds + subtraction~terms.
\end{equation}
Using the relation $<B|\bar{b}i\gamma_{5}u|0>=\frac{m_B^2}{m_b}f_{B}$,
and isolating the contribution of the
lowest B meson from the spectral density, we have
  \begin{equation}
\rho(s)=\delta(s-m_{B}^2)\frac{m_B^2}{m_b}f_{B}F^{B\to \rho}(0)
       +\rho^{h}(s),
\end{equation}
where $\rho^h(s)$ denotes the spectral density of higher resonances and
continuum states $B^h_{p(s)}$, and is approximated as by invoking so-called
quark-hadron duality ansatz
  \begin{equation}
\rho^h(s)=\rho_{QCD}(s)\Theta(s-s_0)
\end{equation}
with the perturbative spectral density $\rho_{QCD}(s)$ and the threshold
parameter $s_0$. Putting everything together, we obtain the hadronic
expression for $F((p+q)^2)$:
  \begin{equation}
F((p+q)^2)=\frac{m^2_B f_B F^{B\to \rho}(0)}{m_b[m_B^2-(p+q)^2]}
+\int_{s_0}^\infty
\frac{\rho_{QCD}(s)}{s-(p+q)^2}ds.
\end{equation}
\par On the other hand, the form factor $F_\mu((p+q)^2)$ is calculable in the
region of large spacelike momenta $(p+q)^2\ll0$ by the OPE near the light
cone $x^2=0$. Because of the large virtuality of the b quark, it is safe 
to calculate the form factor in the light-cone expansion. Here we confine
ourselves to the leading contribution in the OPE, which comes from the
contraction of the b-quark operators to the free b-quark propagator:
   \begin{eqnarray}
<0|Tb(x)\bar{b}(0)|0>&=&\int\frac{d^4p}{(2\pi)^4}e^{-ipx}i\frac{\rlap/p+m_b}
                       {p^2-m_b^2}\\
&=&\int_0^{\infty}\frac{d\alpha}{16\pi^{2}\alpha^2}\nonumber(m_b+i\frac
{\rlap/x}{2\alpha})e^{-m_b^{2}\alpha+\frac{x^2}{4\alpha}}.
\end{eqnarray}
It has been shown in Ref.[4] that the gluon radiative corrections make less
than $5\%$ contribution and can be ignored in our calculation.
Substituting Eq.(14) into Eq.(6), we obtain
  \begin{equation}
F_{\mu}(p,q)=2m_b\int_{0}^{\infty}\frac{d\alpha}
                  {16\pi^{2}\alpha^{2}}
                 \int dxe^{iq\cdot x-m_{b}^2\alpha+\frac{x^2}{4\alpha}}
                 <\rho(p,\eta)|\bar{d}(x)
                 \sigma_{\mu\nu}q^{\nu}(1+\gamma_5)u(0)|0>.
\end{equation}
With the correlator we choose, only the chiral current matrix element
 $<\rho(p,\eta)|\bar{d}(x)\sigma_{\mu\nu}q^{\nu}(1+\gamma_5)u(0)|0>$ remains,
while the chiral-even operator matrix elements
$<\rho|\bar{d}(x)\gamma_{\mu}u(0)|0>$
and $<\rho|\bar{d}(x)\gamma_{\mu}\gamma_{5}u(0)|0>$ disappear.
This will be very useful to improve the sum rule for the form factor $\F$.
If we neglect the corrections from $\rho$-meson mass, 
the matrix element
$<\rho(p,\eta)|\bar{d}(x)\sigma_{\mu\nu}q^{\nu}(1+\gamma_5)u(0)|B>$
can be defined as Ref.[9]
  \begin{equation}
<\rho(p,\eta)|\bar{d}(x)\sigma_{\mu\nu}q^{\nu}(1+\gamma_5)u(0)|0>
            =i[(q\cdot\eta)p_{\mu}\nonumber\\
            -(p\cdot{q})\eta_{\mu}+i\epsilon_{\mu
            \nu\alpha\beta}\nonumber\\
             e^{\nu}q^{\alpha}p^{\beta}]
            \int_{0}^{1}du e^{iup\cdot x}f_\perp \varphi_{\perp}(u),
\end{equation}
where the variable u is the fraction of $\rho$-meson momentum
carried by the d quark and
the value of decay constant $f_\perp$ can be found in Ref.[10],
$f_\perp=(160\pm10)$ MeV; the leading
twist wavefunction $\varphi_\perp(u)$ should be taken at the scale
characterizing virtuality of the 
b quark, $\mu^2 \simeq m^2_B-m^2_b \simeq 5 GeV^2$.
\par At present, the $F((p+q)^2)$ in light-cone QCD sum rule reads
  \begin{equation}
F((p+q)^2)=m_b\int_0^{\infty}\frac{d\alpha}{16\pi^2\alpha}
                            \int dx e^{iq\cdot x-m_b^{2}\alpha
                                +\frac{x^2}{4\alpha}}
             \int_0^{1}du e^{iup\cdot x}\varphi_{\perp}(u,\mu^2) f_\perp.
\end{equation}
Completing the integral over x and $\alpha$, and replacing the variable u by
 $s=\frac{m_b^2}{u}+(1-u)m_\rho^2$ , we have the form factor
  \begin{equation}
F((p+q)^2)=m_b\int_{m_b^2}^{\infty}\frac{\varphi_{\perp}
            (u,\mu^2)}{s-(p+q)^2}
           \frac{u}{u^{2}m_{\rho}^2+m_b^2}f_\perp ds.
\end{equation}
Making the Borel transformation with respect to the variable $(p+q)^2$ in
the hadronic and the QCD expressions and then equating Eq.(13) with Eq.(18), we
obtain the desired sum rule for the form factor $\F$:
  \begin{equation}
\frac{m_B^2}{m_b}f_{B}F^{B\to \rho}(0)e^{\frac{-(m_{B}^{2}-m_{b}^2)}{T}}
=\int_{u(s_0)}^{1}
     \frac{m_{b}f_{\perp}}{u}{\varphi_{\perp}(u,\mu^2)}e^{\frac{u-1}{T}
     (\frac{m_b^2}{u}+m_\rho^2)} du,
\end{equation}
where T is the Borel parameter.
\par It has been shown from the above expression that
the distribution wavefunctions
$\varphi_\parallel$, $g^\nu_\perp$ and $g^a_\perp$ disappear in Eq.(19) and
the uncertainty due to them is avoided. Thus our sum rule for $\F$ to
the twist-3 accuracy is valid, if the corrections from $\rho$-meson mass 
are not taken into account.
\par Now we will choose some input parameters for the sum rule (19), which 
are the light-cone wavefunction $\varphi_\perp(u,\mu^2=5GeV^2)$, decay 
constant $f_B$ and
b-quark pole mass $m_b$. For the same reason as in Ref.[6], we can set
$m_b=4.7, 4.8$ and $4.9 GeV$, accordingly with $f_B=165, 120$ and $85 MeV$. 
For the light-cone wavefunction $\varphi_\perp(u,\mu^2=5GeV^2)$, 
taking into account isospin symmetry we make use of
the same form for $\rho^+$, $\rho^-$, $\rho^0$ and $\omega$:
   \begin{equation}
\varphi_\perp(u,\mu^2)=6u(1-u)[1+0.138\times7.5(\xi^2-0.2)]
   \end{equation}
   with $\xi=2u-1$ in Ref.[10] and
   \begin{equation}
\varphi_\perp(u,\mu^2)=6u(1-u)[1+0.077\times7.5(\xi^2-0.2)-
0.077\times(39.375\xi^4-26.25\xi^2+1.875)]
   \end{equation}
   with $\xi=2u-1$ in Ref.[11] respectively. 
\par A comparison of the wavefunctions, which are respectively 
introduced by P.Ball and V.M.Braun in Ref.[10] and 
A.P.Bakulev and S.V.MIkhailov in Ref.[11], 
is depicted in Fig.1. Numerical calculations corresponding to the 
different wavefunctions will be discussed in the following two cases.
\par Case a, with the light-cone wavefunction Eq.(20), 
 the best stability in the sum rules for $\F$ exists in the
window $T=6.0-10.0 GeV^2$, with a threshold
$s_0=32-34 GeV^2$. Then we obtain the form factors 
$F^{B\to\rho^{\pm}}(0)=0.335\pm0.050$, 
$F^{B\to\rho^0/\omega}(0)=0.237\pm0.035$ and the
branching ratio $Br(B\to\rho^{\pm}\gm)=(2.71\pm1.00)\times10^{-6}$, 
$Br(B\to(\rho^0, \omega)+\gamma)=(1.36\pm0.50)\times10^{-6}$.
The results are greater than these in Ref.[4] slightly.
\par Case b, with the light-cone wavefunction in Ref.[11],
 the best stability exists in the window
$T=6.0-10.0 GeV^2$, with a threshold
$s_0=30-33GeV^2$. This leads the form factor
$F^{B\to\rho^{\pm}}(0)=0.272\pm0.029$, 
$F^{B\to\rho^0/\omega}(0)=0.192\pm0.021$ and the 
branching ratio $Br(B\to\rho^{\pm}\gm)=(1.79\pm0.61)\times10^{-6}$, 
$Br(B\to(\rho^0, \omega)+\gamma)=(0.90\pm0.31)\times10^{-6}$.
This result accords with that $F^{B\to\rho^\pm}(0)=0.28$ in Ref.[4].
\par The variations of $\F$ with the threshold $s_0$
and the Borel parameter T are drawn in Fig.2 and Fig.3.
Here we would like to emphasize that the form factor $\F$ is dependent on
the decay constant $f_B$ and the factor 
$m^2_b \mbox{Exp}({-m^2_b\over{Tu}})$ in Eq.(19).
 This is the reason
why the large variation of $f_B$ with $m_b$ cannot make the form factor $\F$
a great difference. The thresholds $s_0$ in our case (a and b) are both
lower than that in Ref.[4] because the mass of the lowest scalar B meson is
smaller than that of the first excited pseudoscalar B meson. Compared with
the experimental observation in CLEO$\amalg$, 
$Br(B^0\to\rho^0\gm)\leq 3.9\times10^{-5}$,
$Br(B^0\to\omega\gm)\leq 1.3\times10^{-5}$ and
$Br(B^-\to\rho^-\gm)\leq 1.1\times10^{-5}$
in Ref.[1], our results with the two kinds of light-cone wave
functions are below the experimental upper limit.	 
\par In summary, we have reanalyzed the rare process
$B \rightarrow (\rho, \omega)+ \gamma$ in the light-cone
QCD sum rule framework by introducing a chiral current correlator, 
which is different from that in Ref.[4]. 
In this approach, the chiral-even operator matrix
elements, which have not been parametrized precisely by light-cone wave
functions, disappear for the form factor $\F$. 
Further, neglecting the corrections from $\rho$-meson mass,
we have derived the sum rules for form factors to twist-3 accuracy. 
Because of the sensitivity of the numerical results to the decay
constant $f_B$, we have given precision values of the decay constant $f_B$
to leading order in QCD, with the b-quark pole mass
$m_b=4.7-4.9 GeV$. Numerically, the form factors,
$F^{B\to\rho^{\pm}}(0)=0.335\pm0.050$,
$F^{B\to\rho^0/\omega}(0)=0.237\pm0.035$
and the branching ratios
$Br(B\to\rho^{\pm}\gm)=(2.71\pm1.00)\times10^{-6}$,
$Br(B\to(\rho^0, \omega)+\gamma)=(1.36\pm0.50)\times10^{-6}$ 
with wavefunctions of P.Ball and V.M.Braun; 
and the form factors 
$F^{B\to\rho^{\pm}}(0)=0.272\pm0.029$, 
$F^{B\to\rho^0/\omega}(0)=0.192\pm0.021$
and the branching ratios
$Br(B\to\rho^{\pm}\gm)=(1.79\pm0.61)\times10^{-6}$,
$Br(B\to(\rho^0, \omega)+\gamma)=(0.90\pm0.31)\times10^{-6}$
with those of A.P.Bakulev and S.V.Mikhailov. 
Both of them are below the experimental upper limit. 
The different predictions will be tested by future precise experiments.

\newpage
\par
{\large\bf Figure captions}
\par
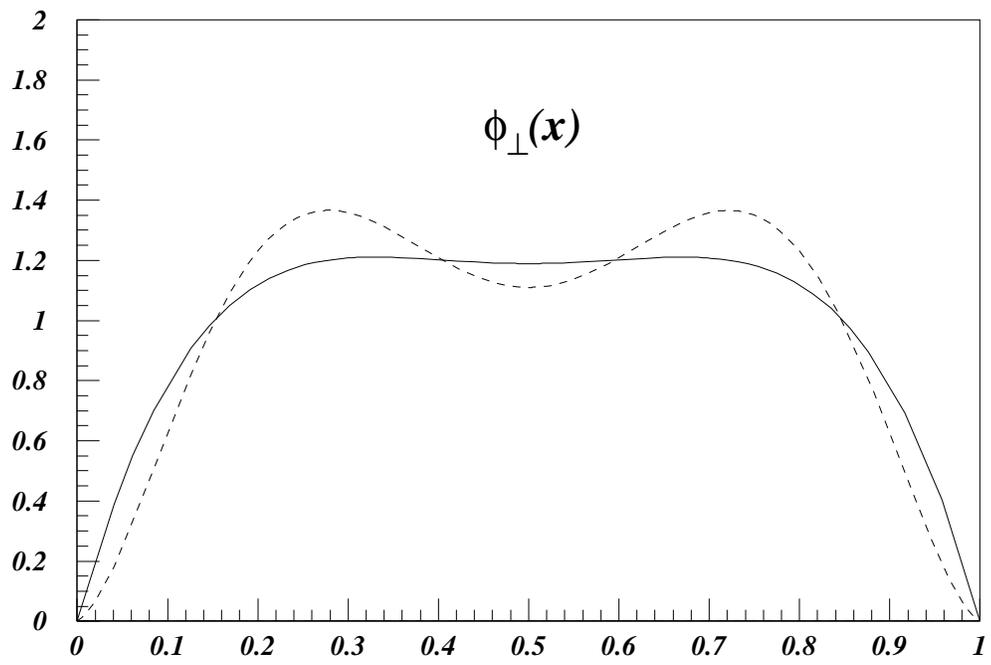
Figure 1: Curves for the different wavefunctions of the $\rho$ meson. 
The solid line represents the one of P.Ball and V.M.Braun, 
and the dashed one represents the one of A.P.Bakulev and S.V.Mikhailov.
\par
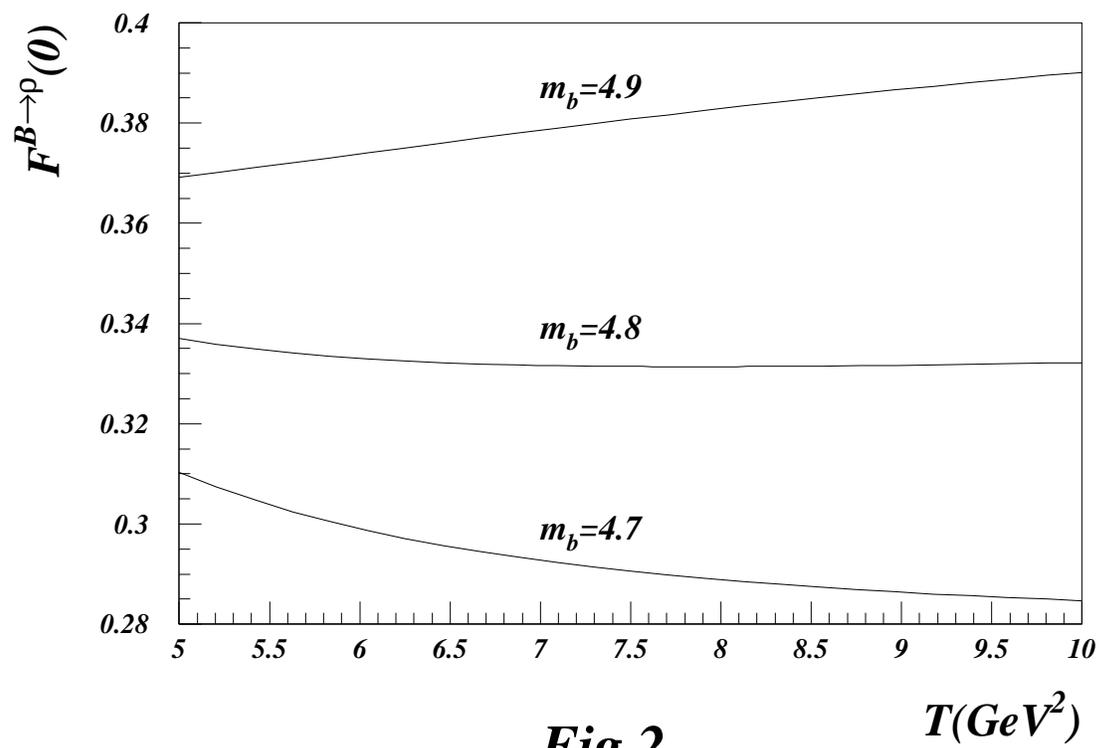
Figure 2: The form factors $\F$ corresponding to the 
threshold $s_0=33 GeV^2$ using the wavefunction of P.Ball and V.M.Braun.
With $m_b=4.7, 4.8$ and $4.9 GeV$ respectively, 
the best fits are shown in the window $T=6-10 GeV^2$.
\par
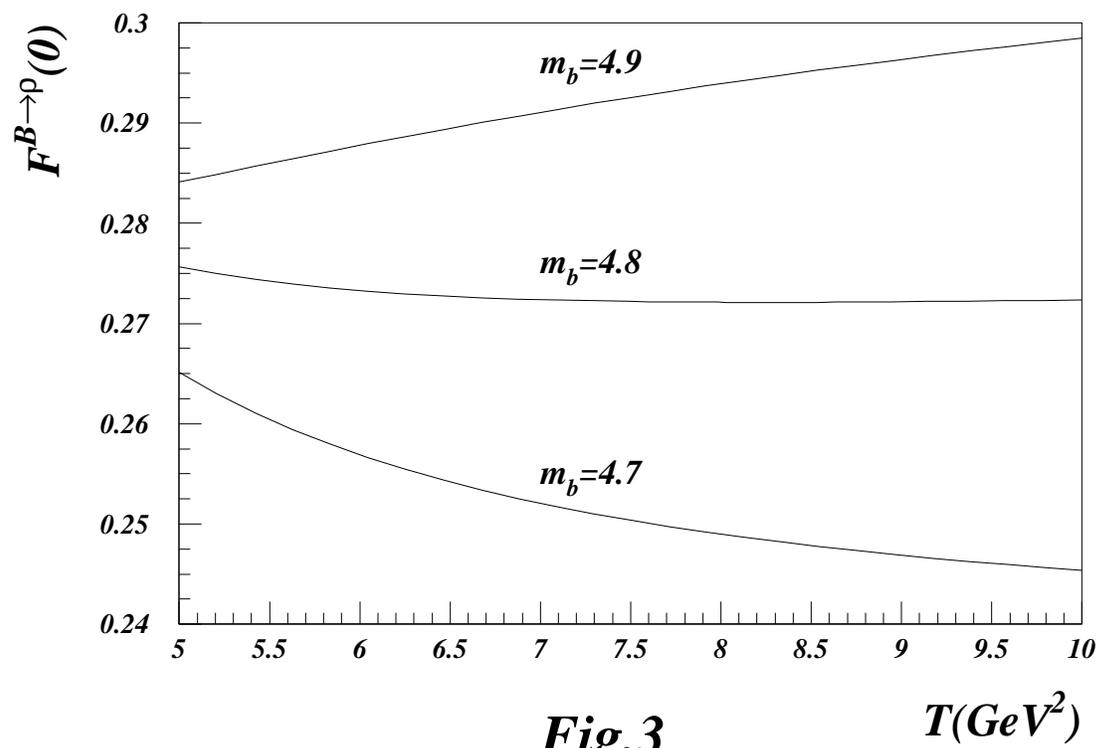
Figure 3: The form factors $\F$ corresponding to the 
threshold $s_0=31.5 GeV^2$ using the wavefunction of 
A.P.Bakulev and S.V.Mikhailov.
With $m_b=4.7, 4.8$ and $4.9 GeV$ respectively, 
the best fits are shown in the window $T=6-10 GeV^2$.


\begin{thebibliography}{99}
\bibitem{1}K.Lingel, T.Skwarnicki and J.G.Smith, hep-ex/9804015.
\bibitem{2} I.I Balitsky et al., Sov.J.Nucl.Phys. {\bf 44} (1986) 1028,
Nucl.Phys. {\bf B 312} (1990) 509;\\ 
V.L.Chernyak and I.R.Zhitnitsky, Nucl.Phys. {\bf B 345} (1990) 137;\\ 
P.Ball, V.M.Braun and H.C.Dosch, Phys.Rev.{\bf D 44} (1991) 3567;\\ 
V.M.Belyaev, A.Khodjamirian and R.Ruckl, Z. Phys {\bf C 60} (1993) 349.
\bibitem{3}P.Ball and V.M.Braun, Phys. Rev. {\bf D 55} (1997) 5561.
\bibitem{4}P.Ball and V.M.Braun, Phys. Rev. {\bf D 58} (1998) 094016.
\bibitem{5}A.Ali, V.M.Braun and H.Simma, Z. Phys. {\bf C 63} (1994) 437.
\bibitem{6}Tao Huang and Zuo-Hong Li, Phys. Rev. D {\bf 57} (1998) 1993.
\bibitem{7}A.Ali, C.Greub, Phys. Lett. B {\bf 287} (1992) 191.
\bibitem{8}B.Grinstein, R.Springer and Mark B.Wise, Nucl. Phys. {\bf B339}
(1990) 269;\\ 
              T.M.Aliev et al., Phys.Lett. {\bf B 237} (1990) 569;\\
              C.A.Dominguez et al., Phys Lett. {\bf B 214} (1988) 459;\\ 
              N.Paver and Riazuddin, Phys.Rev. {\bf D 45} (1992) 978;\\
              S.Narison, Phys.Lett. {\bf B 327} (1994) 354;\\
              P.Colangelo et.al., Phys.Lett. {\bf B 317} (1993) 183.\\
              P.Ball, TU-Munchen Report TUM-T31-43/93.
\bibitem{9} V.L.Chernyak and A.R.Zhitnitsky, Phys.Rep. {\bf 112}
               (1984) 173.
\bibitem{10}P.Ball and V.M.Braun, Phys. Rev. D {\bf 54} (1996) 2182.
\bibitem{11}A.P.Bakulev and S.V.Mikhailov, Phys. Lett. {\bf B 436} (1998) 351.
\end{thebibliography}
\end{document}